\documentstyle[preprint,aps,prl,psfig,floats]{revtex}
\begin{document}

\title{Local Optical Spectroscopy in Quantum Confined Systems:  A
Theoretical Description}

\author{Oskar Mauritz, Guido Goldoni, Fausto Rossi, and Elisa Molinari}
\address{Istituto Nazionale per la Fisica della Materia (INFM),\\
and Dipartimento di Fisica,
Universit{\`a} di Modena, Via Campi 213A, I-41100 Modena, Italy}

\date{\today}

\maketitle
\begin{abstract}

A theoretical description of local absorption is proposed
in order to investigate spectral variations on a length scale
comparable with the extension of the relevant quantum states. A general
formulation is derived within the density-matrix formalism
including Coulomb correlation, and applied to the prototypical case of
coupled quantum wires. The results show that excitonic effects
may have a crucial impact on the local absorption with implications for the
spatial resolution and the interpretation of near-field optical spectra.

\end{abstract}


\narrowtext

The achievement of very high spatial resolutions in optical spectroscopies
of molecules and solids is among the important experimental advancements
of recent years. While in conventional optical experiments the light
field is essentially constant in amplitude and phase over the spatial
extension of the relevant quantum mechanical states, microprobe techniques
make use of highly inhomogeneous light fields.
In the case of low-dimensional semiconductors, diffraction-limited
confocal microscopy has allowed the study of individual
nanostructures and their geometrical fluctuations \cite{confocal}.
With near-field scanning optical microscopy (NSOM), the spatial resolution
is reduced below the diffraction limit and approaches the
scale of quantum confinement \cite{SNOM}: optical spectroscopy thus becomes
a powerful probe of the spatial distribution of quantum states.

In analogy with ultrafast time-resolved spectroscopies \cite{Shah},
that have shown the importance of phase coherence
in the quantum-mechanical time evolution of photoexcited
carriers \cite{Baumberg}, it may be expected that spatial
interference of quantum states play a dominant role
when variations of the electromagnetic (EM-) field occur on an
ultra-short length scale.
On the theoretical side, however, not much work has been done
to investigate the response under these conditions. Most work
has focused on the near-field distribution of the EM-field \cite{chang97}
and its interaction with arrays of point-like particles \cite{hanewinkel97},
and only recently the effect of an
inhomogeneous EM-field on single-particle transitions has been modeled
for a semiconductor quantum dot \cite{bryant98}.

In this Letter we propose a theoretical formulation,
not limited to low photoexcitation densities,
based on a fully microscopic description
of electronic quantum states and their Coulomb correlation.
We show that the non-local character of light-matter interaction
must be taken into account, and demonstrate that
for any given shape of the EM-field distribution
a proper definition of local absorption can still be introduced.
By applying this scheme to the prototypical case of a coupled
quantum-wire structure, we prove that,
when very high-resolution is achieved,
non-local and Coulomb-correlation effects
dominate the local spectra which,
therefore, cannot be simply interpreted as a map of the single-particle
wavefunctions.

The macroscopic polarization ${\bf P}({\bf r},\omega)$
induced by an electromagnetic
field ${\bf E}({\bf r},\omega)$ is in general given by
\begin{equation}\label{macro:pol}
{\bf P}({\bf r},\omega)=
\int \bbox{\chi}({\bf r},{\bf r'},\omega)
\cdot{\bf E}({\bf r'},\omega)\,d{\bf r'} ,
\end{equation}
where $\bbox{\chi}({\bf r},{\bf r'},\omega)$ is the non-local
susceptibility tensor.
It can be regarded as the Fourier transform of the local polarization,
whose microscopic expression is given by
\begin{equation}\label{micro:pol-1}
{\bf P}({\bf r},t)=q\left\langle {\bf \hat{\Psi}}^\dagger({\bf r},t){\bf r}
{\bf \hat{\Psi}}({\bf r},t) \right\rangle,
\end{equation}
where $q$ is the electronic charge,
$\left\langle \ldots \right\rangle$ denotes
a proper ensemble average
and the field operator ${\bf \hat{\Psi}}({\bf r},t)$ in the Heisenberg
picture describes the microscopic evolution of the carrier system.
Within the usual electron/hole picture,
the optical (i.e., interband) contribution to
${\bf P}({\bf r},t)$ can be expressed in terms
of the non-diagonal elements of the single-particle density matrix~~
$p_{eh}=\langle \hat{d}_h \hat{c}_e\rangle$
($\hat{d}_h$ and $\hat{c}_e$ being
the destruction operators for a hole in state $h$ and an electron
in state $e$) according to:
\begin{equation}\label{micro:pol-2}
{\bf P}({\bf r},t)= q \sum_{eh} \left[p_{eh}(t)
\Psi_e({\bf r}){\bf r} \Psi_h({\bf r}) +\mbox{c.c.}\right]\ .
\end{equation}
Here, $e$ and $h$ are appropriate sets of quantum numbers which label
the free-carrier wavefunctions $\Psi_{e/h}$ involved in
the optical transition.

The time evolution of the single-particle density matrix
is described by the Semiconductor Bloch Equations \cite{Kuhn},
which for the non-diagonal elements (interband polarizations) read
\cite{rossi96}:
\begin{equation}
\frac{\partial p_{eh}}{\partial t} =
\frac{1}{i\hbar}({\cal E}_e+{\cal E}_h)p_{eh}+
\frac{1}{i\hbar}{\cal U}_{eh}(1-f_e-f_h)+
\frac{\partial p_{eh}}{\partial t} \Bigg\vert_{ \mbox{\scriptsize coll}}\, ,
\label{dpdt}
\end{equation}
where ${\cal U}_{eh}$ and ${\cal E}_{e/h}$ are respectively the
EM-field and single-particle energies renormalized by the
Coulomb interaction.
The last (collision) term in Eq.\ (\ref{dpdt})
accounts for incoherent (i.e., scattering and diffusion)
processes \cite{elsasser}.
Stationary solutions of this equation can be found by assuming
equilibrium distribution functions $f_e$, $f_h$;
in this case Eq.\ (\ref{dpdt}) can be
transformed into an eigenvalue equation whose $\lambda$-th solution
provides the complex energy eigenvalue
$\hbar\omega^\lambda = \epsilon^\lambda-i\Gamma^\lambda$ and
eigenvector $p^\lambda_{eh}$.

By inserting the stationary solution
$p_{eh}(t) = p^\lambda_{eh} e^{-i\omega^\lambda t}$ of
Eq.\ (\ref{dpdt}) into Eq.\ (\ref{micro:pol-2}) and Fourier transforming,
it is possible to express the polarization ${\bf P}({\bf r},\omega)$
in the form of Eq.\ (\ref{macro:pol}), from which a microscopic expression for
$\bbox{\chi}$ can be identified.
In particular, for semiconductor structures described within the
usual envelope-function formalism with isotropic electron and hole
bulk dispersions,
the susceptibility tensor $\bbox{\chi}$
becomes diagonal, with identical elements given by
\begin{eqnarray}\label{chi}
&&\chi({\bf r},{\bf r'},\omega)=\vert M_b \vert^2 \times\\
\nonumber &&\sum_{\lambda,eh,e'h'}p^\lambda_{eh} p^{\lambda *}_{e'h'}
(1-f_{e'}-f_{h'})
\frac{\psi_e({\bf r})\psi_h({\bf r})
\psi^{*}_{e'}({\bf r'})\psi^{*}_{h'}({\bf r'})}{
\epsilon^\lambda-\hbar\omega-i\Gamma^\lambda}.
\end{eqnarray}
Here, $\psi_{e/h}({\bf r})$ are single-particle electron/hole envelope
functions and $M_b$ is the bulk optical matrix element.

In the usual definition of the absorption coefficient
within the dipole approximation, the non-locality of $\chi$ is neglected.
When non-locality is taken into account, it is no longer possible to
define an absorption coefficient that locally relates the absorbed
power density with the light intensity. However, considering a light
field with a given profile $\xi$ centered around the beam position
${\bf R}$, $E({\bf r},\omega)=E(\omega)\xi({\bf r}-{\bf R})$,
we may define a local absorption that is a function of the
beam position, and relates the {\em total}
absorbed power to the power of a {\em local} excitation
(illumination mode):
\begin{equation}\label{alp}
\alpha_\xi({\bf R},\omega)
\propto\mbox{\hspace{-2truemm}}\int\mbox{\hspace{-2truemm}}\;\Im 
\left[  \chi({\bf r},{\bf r'},\omega)
\right] \xi({\bf r}-{\bf R})\xi({\bf r'}-{\bf R})\,d{\bf r}\,d{\bf r'}.
\nonumber
\end{equation}
This expression is in principle not limited to low-photoexcitation
intensities; together with Eq.\ (\ref{chi}) it provides a general
description of linear as well as non-linear local response, i.e., from
excitonic absorption to the gain regime.

In the linear-response regime $1-f_e-f_h\simeq1$ and the quantity
$\Psi^\lambda({\bf r_e},{\bf r_h})=
\sum_{eh} p^\lambda_{eh}\psi_e({\bf r_e})\psi_h({\bf r_h})$
can be identified with the exciton wavefunction; in this case
the non-local susceptibility (\ref{chi}) reduces to the result of
linear-response theory as given, e.g., in \cite{dandrea82}.
In this regime the contribution of the $\lambda$-th excitonic state
to the local spectrum is
\begin{equation}\label{exc_contrib}
\alpha_\xi({\bf R},\omega^\lambda)
\propto
\left\vert \int \Psi^\lambda({\bf r},{\bf r})\xi({\bf r}-{\bf R})
\,d{\bf r}\, \right\vert^2.
\end{equation}

The above formulation is valid for semiconductors of
arbitrary dimensionality.
To illustrate the effects of non-locality and Coulomb correlation
on the local absorption spectrum, we now consider
quasi-one-dimensional (1D) nanostructures
(quantum wires), subject to a local EM excitation propagating
parallel to the free axis of the structure, $z$.
For simplicity, we describe the narrow light beam by a gaussian
EM-field profile,
$\xi({\bf r}) = \exp[-(x^2+y^2)/2\sigma^2]$
\cite{EM-field}.
For a quantum wire the single-particle electron/hole wave-functions
appearing
in (\ref{chi}) can be written as $\psi_e({\bf r})=\phi_{\nu_e}(x,y)
e^{i k^e_z z}$, $\psi_h({\bf r})=\phi_{\nu_h}(x,y) e^{i k^h_z z}$
where $\nu_{e/h}$ are subband indices and $k^{e/h}_z$ are wavevectors
along the free axis.

\begin{figure}
\noindent
\unitlength1mm
\begin{picture}(100,80) 
\put(20,5){\psfig{figure=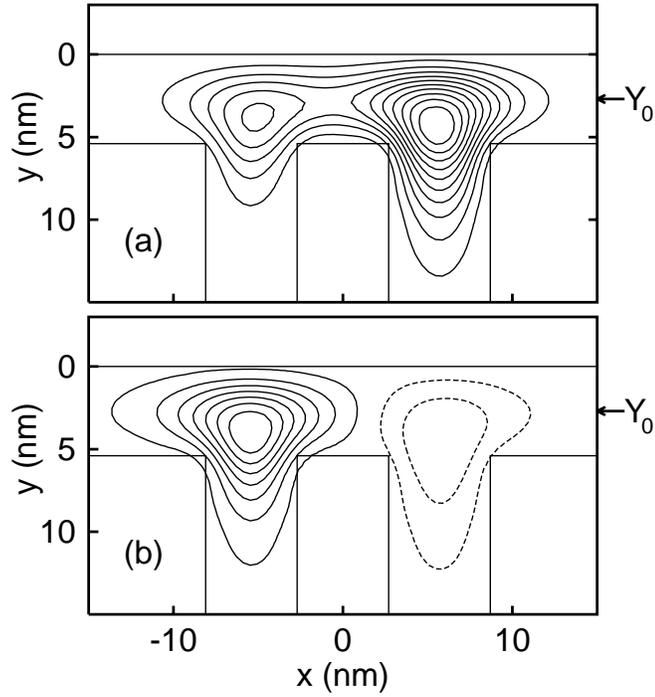,width=110mm}}
\end{picture}
\caption{
Effective exciton wavefunctions $\Phi^{\lambda}(x,y)$
for two coupled T-shaped GaAs/AlAs quantum wires,
formed at the intersections between two vertical quantum wells
(5.4 and 6.0 nm wide, with a 5.4 nm barrier) and a horizontal
quantum well (5.4 nm wide). Only the dominant (real) part
of the complex wavefunction is shown for
{(a)} the ground state exciton,
and {(b)} one of the excited states giving rise to
the shoulder labelled D
in Figs.\ 2 and 3. Dashed contours correspond to
negative values. Note the two distinct regions
with different sign, giving rise to cancellation effects for
sufficiently large values of $\sigma$.}
\label{Fig:wavefun}
\end{figure}

To discuss the 1D case it is convenient to take advantage of the
translational invariance along $z$ and define
\begin{eqnarray}
\Phi^{\lambda}(x,y) & \equiv & \int{\Psi}^\lambda({\bf r},{\bf r})\,dz
\nonumber\\
&\propto& \sum_{\nu_e\nu_h k_z} p^\lambda_{\nu_e k_z \nu_h -k_z}
\phi_{\nu_e}(x,y)\phi_{\nu_h}(x,y).
\end{eqnarray}
We shall refer to $\Phi^{\lambda}(x,y)$ as the {\em effective}
exciton wavefunction.
$\Phi^{\lambda}(x,y)$ enters Eq.\ (\ref{exc_contrib}) and, convoluted
with the spatial distribution of the EM-field, $\xi$, yields the
contribution of the $\lambda$-th excitonic state
to the local absorption $\alpha_\xi(X,Y,\omega)$;
note that only Fourier components of the polarization with
$k^e_z=-k^h_z$ contribute to the absorption.

The effects of spatial coherence of quantum states
are understood most easily in the linear regime on the basis of
Eq.\ (\ref{exc_contrib}). For a spatially  homogeneous
EM-field, the absorption spectrum probes the average
of $\Phi^{\lambda}$ over the whole space (global spectrum).
In the opposite limit of an infinitely narrow
probe beam, $\alpha(X,Y,\omega^\lambda)$ maps
$\vert \Phi^{\lambda}(X,Y)\vert^2$; the local absorption is non-zero
at any point where the effective wavefunction of an exciton gives a finite
contribution.
It is therefore clear that 'forbidden' excitonic transitions,
not present in the global spectrum, may appear in the local one.
In the intermediate regime of a narrow but finite probe, it is possible
that a cancellation of the contributions from $\Phi^{\lambda}(x,y)$
takes place between different points in space leading to a non trivial
localization of the absorption. The result will then be quite sensitive
to the extension of the light beam.

As a prototype system showing non-local effects we
consider two coupled GaAs/AlAs T-shaped wires \cite{TWIRES}. To obtain
spatially separated transitions we consider an asymmetric structure
(Fig.\ 1); the barrier width is chosen to allow Coulomb coupling between
the wires \cite{3D}.
For both electrons and holes, the single-particle ground state (E1, H1)
is localized in the widest wire (right wire, RW) and
the second bound state (E2, H2) in the left wire (LW) \cite{holes}.
The effective exciton wavefunctions $\Phi^{\lambda}$
are shown in Fig.\ 1 for the ground state exciton
and for one of the excited states.
They are found to differ significantly from the single particle
wavefunctions (not shown here); this is true in general for
coupled nanostructures.
The ground state exciton mainly involves single-particle
contributions from the first electron and first hole
subband (E1-H1), and has no nodes; hence, no cancellation is possible
and its contribution to the local spectrum
is expected to be finite for any extension of the probe beam.
On the other hand the effective exciton wavefunction for the higher state
has areas with different signs; therefore its convolution with
$\xi$ may produce cancellations, whose results depend strongly on
the characteristic spatial extension of the beam, $\sigma$.
Of course, the phase of the relevant quantum states is determined by
electron-hole correlations on the scale of the exciton Bohr radius;
therefore, to emphasize these effects, in the following we show calculations
for a beam with $\sigma=$ 10 nm (close to the Bohr radius 
$a_\circ$ in GaAs). Note that NSOM experiments on semiconductors
are currently still limited to higher values of $\sigma$;
calculations performed for  $\sigma \gg a_\circ$ confirm that 
these effects become negligible, so that a local description
is sufficient. 

\begin{figure}
\noindent
\unitlength1mm
\begin{picture}(100,110) 
\put(30,-10){\psfig{figure=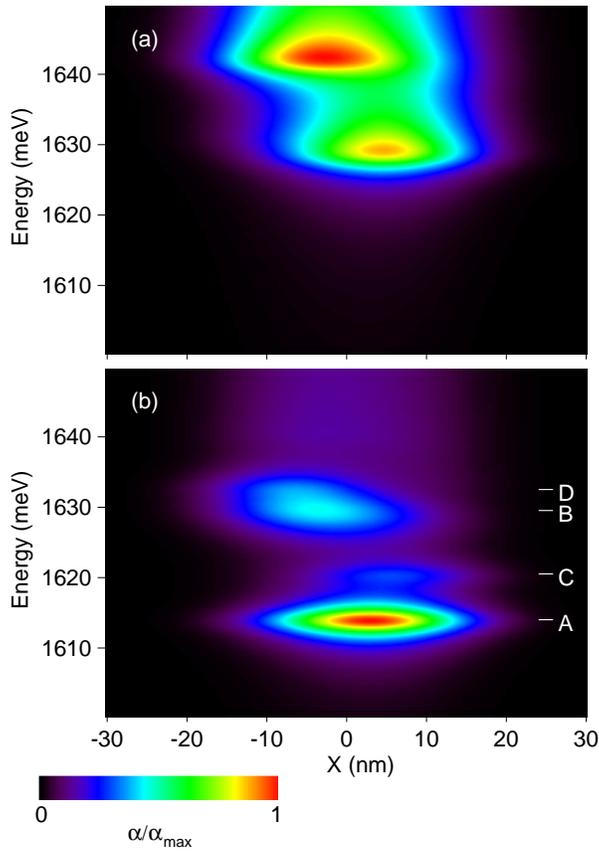,width=90mm}}
\end{picture}
\caption{
Local absorption $\alpha(X,Y,\hbar\omega)$ as a function of photon energy
and beam position. The beam coordinate $X$ varies across the wires
parallel to the $x$ direction, with $Y=Y_0$ fixed at the center of
the horizontal quantum well (see Fig.\ 1); $\sigma=10$ nm.
The absorption is normalized so
that the maximum absorption, $\alpha_{\mbox{max}}$ is the same in both maps.
(a) Spectrum calculated in the single-particle
approximation; (b) full calculation, with the electron-hole
Coulomb correlation taken
into account. The labels A, B, C, D identify
the main structures discussed in the text.}
\label{Fig:xEmap}
\end{figure}

Figure 2 shows the local absorption
$\alpha(X,Y,\hbar\omega)$ in the linear regime
as a function of photon energy
and beam position. $Y=Y_0$ is fixed at the center of
the horizontal quantum well (see Fig.\ 1), while the $X$ coordinate varies
across the structure. As a reference we first discuss results
in the absence of electron-hole interaction [Fig.\ 2(a)]:
the spectrum is essentially
composed of two structures, both with the typical inverse-square
root high energy tail ensuing from 
the 1D free-particle density of states.
The signal from the E1-H1 transition is spatially located on the
RW while a second peak, located on the LW, stems
from the E2-H2 transition; in this uncorrelated case the
influence of spatially indirect transitions (E1-H2, E2-H1) is negligible.
The correlated carrier spectrum [Fig.\ 2(b)] is very
different. As expected, the spectrum is dominated
by excitonic peaks at lower energies (the ground state binding
energy is 14 meV \cite{Rossi97}),
and the corresponding continuum is strongly suppressed
by the electron-hole interaction \cite{rossi96}.
The two main peaks (A and B, at $\approx$\ 1614 and 1629 meV,
respectively) still have their largest contributions in the RW and LW,
respectively; however, two weaker structures C and D appear
(at $\approx$\ 1620 meV and $\approx$\ 1633 meV, respectively), which
are more strongly localized in either wire and have
no equivalent in the uncorrelated spectrum.

\begin{figure}
\noindent
\unitlength1mm
\begin{picture}(100,90) 
\put(30,10){\psfig{figure=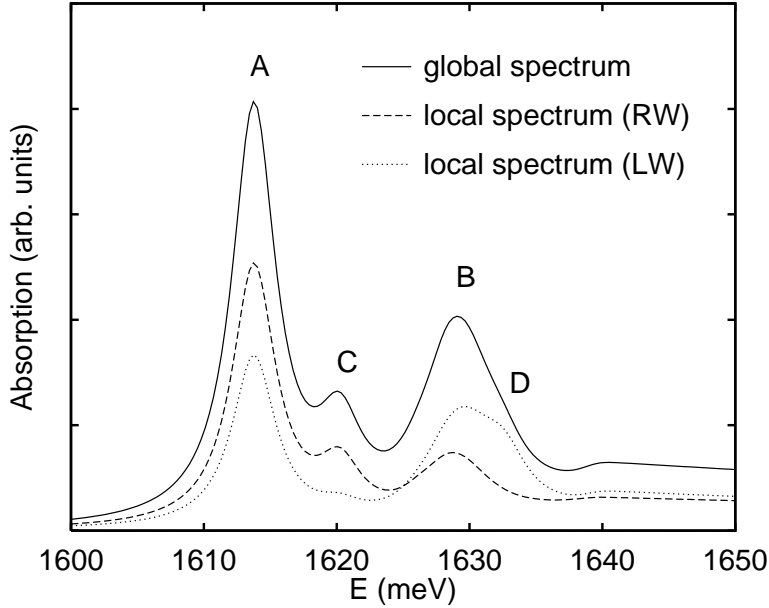,width=100mm}}
\end{picture}
\caption{
Global (solid line) and local absorption spectra
(including electron-hole correlation), calculated
with the beam centered on the right wire (dashed line)
and on the left wire (dotted line); $\sigma=10$ nm.
The labels A, B, C, D identify the main structures
discussed in the text.
}
\label{Fig:spectra}
\end{figure}

We analyse the origin and spatial localization of these structures
with the help of Fig.\ 3, which
shows the local spectra with the beam centered on the
RW or the LW, as well as the global spectrum \cite{cut}.
The most intense peak (A) is the ground state exciton (E1-H1);
it has the strongest contribution from the RW, but a
significant intensity is found also in the LW, consistently
with the spatial extension of its effective wavefunction [Fig.\ 1(a)].
Peak C also
originates from the RW where a second bound exciton is
introduced in the presence of Coulomb coupling with the LW;
indeed, this structure is not present when the wires
are far apart. Peak B stems from several exciton states,
with a major component in the E2-H2 transition on the LW.
The shoulder D (mostly a E2-H1 transition)
is very intense in the local spectrum centered on the LW,
completely absent when the beam is centered on the RW,
and obviously very reduced in the global spectrum.
A detailed analysis shows that the strong localization of
D comes from the interference between positive and negative
regions of the effective exciton wavefunction
[Fig.\ 1(b)], whose cancellation depends
on the position of the beam.

Similar calculations for other types of coupled nanostructures
confirm that these phenomena are general, and especially
significant when they are investigated with spatial resolution 
of the order of the Bohr radius. At such resolutions, 
the breaking of selection rules that hold in integrated spectra 
might then give insight into Coulomb correlation and coherence phenomena.

In summary, we have presented a general formulation of local optical
absorption, holding for any given profile of the light
beam even when its width becomes comparable to the exciton Bohr radius in
the semiconductor. The key ingredient is the effective exciton
wavefunction, whose interference effects have a very different
influence on local and global spectra. As the spatial resolution
of near-field optical experiments increases, these effects
are expected to become important in the interpretation of local spectra.

This work was supported in part by the EC under the TMR Network
"Ultrafast Quantum Optoelectronics".



\begin{references}
\bibitem{EM-field}
The actual profile of the EM field generated by a NSOM tip
is more complex, and a research topic in itself 
\protect\cite{chang97,hanewinkel97,bryant98}.
However, our simple choice for $\xi$ is sufficient to highlight
the effects of nonlocality. An arbitrary field distribution
can be included in our treatment through Eq.\ (\ref{alp}), while 
Eq.\ (\ref{chi}) is independent of the external field.
\bibitem{3D} Calculations are performed according to
Refs.~\onlinecite{rossi96,Rossi97};
the single-particle states are solutions of the
Schr{\"o}dinger-like equation for envelope functions.
The full three-dimensional nature of the
Coulomb interaction is retained by expanding the excitonic state
on the basis of single-particle transitions.
\bibitem{cut} For clarity, in the present calculations we have neglected
higher-energy
states from the wider QW. We have checked that these do not affect the
local spectra
in Fig.\ 3, but would only contribute to the high-energy part
of the global spectrum in the same figure.
\end{references}
\end{document}